\documentclass[11pt,draftcls,onecolumn]{IEEEtran}

\usepackage{etex}
\usepackage{bbm}
\usepackage{balance}
\usepackage{cite}
\usepackage{soul}
\usepackage{graphicx}
\usepackage{epstopdf}
\usepackage{amssymb}
\usepackage[cmex10]{amsmath}
\usepackage{cases}
\let\labelindent\relax
\usepackage{enumitem}
\usepackage{array}
\usepackage{multirow}
\usepackage{bigdelim}
\usepackage{mdwtab}
\usepackage{tikz}
\usepackage{physics}
\usepackage{color}
\usepackage{booktabs}

\makeatletter
\let\MYcaption\@makecaption
\makeatother
\usepackage[font=footnotesize]{subcaption}
\makeatletter
\let\@makecaption\MYcaption
\makeatother

\usepackage{accents}

\newtheorem{theorem}{Theorem}[section]
\newtheorem{lemma}{Lemma}[section]

\title{Entropy-Regularized Partially Observed\\ Markov Decision Processes}%
\author{Timothy L.\ Molloy, \IEEEmembership{Member, IEEE}, and Girish N.\ Nair, \IEEEmembership{Fellow, IEEE}%
\thanks{The first author was with the Dept.\ of Electrical and Electronic Engineering, University of Melbourne, VIC 3010, Australia. He is now with the CIICADA Lab, School of Engineering, The Australian National University (ANU), Canberra, ACT 0200, Australia (e-mail: timothy.molloy@anu.edu.au) The second author is with the Dept.\ of Electrical and Electronic Engineering, University of Melbourne, VIC 3010, Australia (e-mail: gnair@unimelb.edu.au)}
\thanks{This work received funding from the Australian Government, via grant AUSMURIB000001 associated with ONR MURI grant N00014-19-1-2571.}%%
\thanks{Preliminary versions of some results in this paper were presented at the 2022 American Control Conference \cite{Molloy2022}.}%
}

\begin{document}

% ACC
% - propose joint entropy for finite horizon problems
% - establish additive form
% - POMDP and belief MDP reformulations with optimality of deterministic policies
% - uncertainty aware navigation and controlled sensing examples

% Journal version
% - finite potentially randomized
% - conditional and input-output entropy (minimise)
% - additive forms of all entropies (causal entropies)
% - proof deterministic policies only required
% - belief state formulations
% - structural results concave
% - pwlc approximations
% - transmission example
%   - transmission rate
% - joint special case
%   - linear in special case
%   - infinite horizon
%   - sarsop navigation example
%   - active hypothesis testing/greedy

% If the state and observation spaces correspond (i.e., $\mathcal{X} = \mathcal{Y}$), and $B^{x,y}(u) = 1$ if $x = y$ and $B^{x,y}(u) = 0$ if $x \neq y$ for all $u \in \mathcal{U}$ and deterministic policies
% \begin{align*}
%     H_\mu(X^T, Y^T, U^{T-1})
%     &= H_\mu(X^T)
% \end{align*}

% Our JTERPOMDP problem then specializes to the (joint) trajectory-entropy-regularized MDP (JTER-MDP):
% \begin{align}
%     \label{eq:fh_problem}
%     \begin{aligned}
%     &\inf_{\mu \in \mathcal{P}} & & J_{\mu}^T + \beta H_\mu(X^T)\\ 
%     &\mathrm{s.t.} & &  X_{k+1} | X_k, U_k \sim A^{x_{k+1},x_k}(u_k)\\
%     & & & U_k = \mu_k^{i_k} \in \mathcal{U}\\
%     & & & X_0 \sim \rho
%     \end{aligned}
%     \end{align}
% for a given scalar constant $\beta \geq 0$ and a given finite horizon $0 < T < \infty$.

\maketitle
\thispagestyle{empty}

\begin{abstract}
\boldmath
We investigate partially observed Markov decision processes (POMDPs) with cost functions regularized by entropy terms describing state, observation, and control uncertainty.
Standard POMDP techniques are shown to offer bounded-error solutions to these entropy-regularized POMDPs, with exact solutions possible when the regularization involves the joint entropy of the state, observation, and control trajectories. 
Our joint-entropy result is particularly surprising since it constitutes a novel, tractable formulation of active state estimation.
\end{abstract}

% \fi
\IEEEpeerreviewmaketitle

\section{Introduction}
Partially observed Markov decision processes (POMDPs) and Markov decision processes (MDPs) with information-theoretic costs have attracted widespread attention across systems and control \cite{Tanaka2021,Shafieepoorfard2016,Hoffmann2010,Molloy2021a}, computer science \cite{Araya2010,Fehr2018,Rubin2012}, signal processing \cite{Krishnamurthy2007,Krishnamurthy2016, Zois2017, Zois2014}, and robotics \cite{Thrun2005,Stachniss2005,Valencia2012}.
Interest in such POMDPs has been driven, in large part, by \emph{active state estimation} problems in which information-theoretic costs describing the uncertainty about latent states are minimized in order to aid or enhance the performance of state estimation algorithms \cite{Krishnamurthy2007,Krishnamurthy2016,Araya2010,Molloy2021a}.
Interest in such MDPs has, in contrast, been driven by a desire within applications such as networked control and economics to develop control policies (or decision-makers) that are \emph{rationally inattentive} or ``data-frugal'' in that they trade-off control performance to reduce the (data) rate at which state information is used to make control decisions \cite{Tanaka2021,Shafieepoorfard2016,Rubin2012}.
Despite interest in rate-cost trade-offs in MDPs, limited attention has been paid to similar problems in POMDPs.
Motivated by data-frugal POMDPs with potential applications to active state estimation, we investigate POMDPs with information-theoretic entropy costs that penalize observation incompressibility and/or state uncertainty.

POMDPs with information-theoretic costs have been extensively investigated for active state estimation, with popular costs including the (negative) mutual information between states and observations \cite{Hoffmann2010}, the (Shannon or R\'enyi) entropy of Bayesian filter estimates \cite{Krishnamurthy2007,Thrun2005,Araya2010}, and the entropy of Bayesian smoother or Viterbi algorithm estimates \cite{Stachniss2005,Valencia2012,Molloy2021a} (see \cite[Chapter 8]{Krishnamurthy2016} and references therein for more).
These POMDPs have been shown to be amenable to bounded-error (approximate) solution using standard POMDP solvers when their cost and cost-to-go (or value functions) are concave and admit piecewise-linear concave (PWLC) approximations (cf.\ \cite[Section 8.4.4]{Krishnamurthy2016}, \cite{Araya2010}).
For example, we recently showed that the \emph{smoother entropy} (i.e., the conditional entropy of the state trajectory given observations and controls) can be (approximately) optimized in this manner \cite{Molloy2021a,Molloy2021b}.

Outside of applications involving active state estimation, POMDPs with information-theoretic costs have received only limited specialized attention.
Most notably, in Bayesian experimental design (involving degenerate POMDPs with a time-invariant or constant state), the entropy of the observations has been explored as a cost to encourage the selection of controls (i.e. experiments) with predictable outcomes (see \cite{Ryan2016,Chaloner1995}).
In the context of linear-quadratic regulators and linear-quadratic-Gaussian control (i.e.\ POMDPs with specialized linear dynamics but continuous state, control, and observation spaces), the \emph{directed information} from the observations to controls has been used as a cost to study the trade-off between feedback (data) rates and control costs (see \cite{Tanaka2018,Kostina2019,Sabag2020}).
Similarly, in the context of MDPs (i.e.\ degenerate POMDPs with fully observed states), various information-theoretic quantities such as the directed information, mutual information, and \emph{transfer entropy}, have been used as costs to penalize feedback from the states to the controls so as to study rate-cost trade-offs \cite{Tanaka2021,Shafieepoorfard2016,Rubin2012}.
Despite growing interest, solving POMDPs (and MDPs) with rate-cost trade-offs has proved difficult due to complications including randomized policies \cite{Tanaka2021,Tanaka2018,Ryan2016,Chaloner1995}, the design of observation processes \cite{Tanaka2021,Tanaka2018,Shafieepoorfard2016,Rubin2012}, and the need to solve nonconvex optimization problems \cite{Tanaka2021,Fehr2018}.

The main contribution of this paper is the proposal of POMDPs with costs regularized by combinations of the \emph{input-output entropy} (i.e.\ the entropy of the observations and controls) and the smoother entropy.
Such \emph{entropy-regularized} POMDPs (ERPOMDPs) are novel in that they both introduce rate-cost trade-offs into standard POMDPs (due to the relevance of the input-output entropy to rate-cost trade-offs via Shannon's source coding theorem), and generalize recent work on active state estimation involving the smoother entropy to include compressible (or predictable) observations (cf.\ \cite{Molloy2021a,Molloy2021b}).
Importantly, we show that ERPOMDPs admit bounded-error PWLC solutions via standard POMDP techniques, in general, and \emph{exact} solutions \emph{without} additional PWLC approximations in the special case where the smoother and input-output entropies are equally weighted and become the joint entropy of the states, observations, and controls.
The solution of ERPOMDPs involving the joint entropy without PWLC approximations is surprising since the vast majority of other POMDPs with information-theoretic costs are entirely intractable without them.
Compared to our preliminary work in \cite{Molloy2022}, significant extensions in this paper include: 1) consideration of a generalized problem \eqref{eq:er} with arbitrary combinations of the smoother and input-output entropies; 2) development of entirely new results in Lemmas \ref{lemma:causal_entropy} and \ref{lemma:simplified_er}, and Theorem \ref{theorem:MDP}  concerning the input-output entropy; and, 3) new operational interpretations of ERPOMDPs in Section \ref{sec:interp}.

% amenable to direct optimization with standard POMDP solvers 
% Key to the tractability of the joint entropy in POMDPs is that it admits a closed-form algebraic expression, which surprisingly enables us to show that optimizing it involves a search only over deterministic control policies with cost functions that are linear in the belief state.
% Despite similar combinations of conditional and input-output entropies (and the joint-entropy) have previously been explored for optimal Bayesian experimental design for (static) parameter estimation (cf.\ \cite{Chaloner1995,Ryan2016} and references therein), their consideration as an objective in active state estimation is novel.
% Intuitively, minimizing the joint entropy reduces both the uncertainty associated with the state trajectory (making it easier to infer), and the randomness of the control and observation trajectories (reducing the complexity of predicting, storing, and/or communicating them).
%Operationally, minimizing the joint entropy also corresponds to minimizing an upper bound on the probability of error for any state trajectory estimator (via bounding the conditional entropy of the states given the observations and controls, cf.\ \cite{Molloy2021a} and \cite{Feder1994}).

This paper is structured as follows.
In Section \ref{sec:problem}, we pose ERPOMDPs and examine their solution via standard POMDP techniques in Section \ref{sec:reforms}.
In Section \ref{sec:jer}, we examine \emph{exact} ERPOMDP solutions in the case of joint-entropy regularization.
Finally, we provide interpretations of ERPOMDPs in Section \ref{sec:interp}, simulations in Section \ref{sec:results}, and conclusions in Section \ref{sec:conclusion}.

\emph{Notation:}
Random variables will be denoted by capital letters (e.g., $X$), their realizations by lower case letters (e.g., $x$), and associated sequences by letters with superscripts denoting their final times (e.g., $X^T \triangleq \{X_0, X_1, \ldots, X_T\}$ and $x^T \triangleq \{x_0, x_1, \ldots, x_T\}$).
The probability mass function (pmf) of $X$ will be written $p(x)$, the joint pmf of $X$ and $Y$ written $p(x, y)$, and the conditional pmf of $X$ given $Y = y$ written $p(x|y)$ or $p(x | Y = y)$.
For a function $f$ of $X$, the expectation of $f$ is $E_X [f(X)]$ and the conditional expectation of $f$ under $p(x|y)$ is $E[f(X) | y]$.
The \emph{pointwise} conditional entropy of $X$ given $y$ is $H(X | y) \triangleq - E[\log p(X | Y = y) | y]$, and the conditional entropy of $X$ given $Y$ is $H(X|Y) \triangleq E_Y [H(X|y)]$, with the base of the logarithm being 2.
% The mutual information between $X$ and $Y$ is $I(X; Y) \triangleq H(X) - H(X | Y) = H(Y) - H(Y | X)$. 
% The pointwise conditional mutual information of $X$ and $Y$ given $Z = z$ is $I(X; Y | z) \triangleq H(X | z) - H(X | Y, z)$ with the (average) conditional mutual information given by $I(X; Y | Z) \triangleq E_{Z} \left[ I(X; Y | z) \right]$.

\section{Problem Formulation}
\label{sec:problem}

Let $X_k$ for $k \geq 0$ be a discrete-time first-order Markov chain with the finite state space $\mathcal{X} \triangleq \{1, 2, \ldots, N_x\}$.
Let the initial state $X_0$ be distributed according to the pmf $\rho \in \Delta$ with components $\rho(x_0) \triangleq P(X_0 = x_0)$ where $\Delta \triangleq \{\rho \in [0,1]^{N_x} : \sum_{x \in \mathcal{X}} \rho(x) = 1\}$ is the { $(N_x-1)$--dimensional probability simplex}.
Let the (controlled) transition dynamics of the state $X_k$ be described by the transition kernel:
\begin{align}
    \label{eq:stateProcess}
    A^{x,\bar{x}}(u) \triangleq p( X_{k+1} = x | X_k = \bar{x}, U_k = u)
\end{align}
for $k \geq 0$ with the controls $U_k = u$ belonging to the finite set $\mathcal{U} \triangleq \{1, 2, \ldots, N_u\}$.
The state process $X_k$ is (partially) observed through a stochastic observation process $Y_k$ for $k \geq 0$ taking values in the finite set $\mathcal{Y} \triangleq \{1, 2, \ldots, N_y\}$.
The observations $Y_k$ are distributed according to the kernel:
\begin{align}
    \label{eq:obsProcess}
    B^{x,y} (u) \triangleq p( Y_k = y | X_k = x, U_{k-1} = u)
\end{align}
for $k > 0$ with $B^{x_0,y_0} \triangleq p( Y_0 = y_0 | X_0 = x_0)$.
The controls $U_k$ for $k \geq 0$ arise from a potentially stochastic output-feedback policy $\mu \triangleq \{ \mu_k^{i_k} : k \geq 0\}$ with (conditional) pmfs
\begin{align*}
    \mu_k^{i_k}(u_k) 
    \triangleq p(U_k = u_k | Y^k = y^k, U^{k-1} = u^{k-1})
\end{align*}
where $i_k \triangleq (y^{k},u^{k-1})$ is a realization of the \emph{information state} $I_k \triangleq (Y^k, U^{k-1})$.
The joint pmf of $(X^T, I_T)$ under $\mu$ is
\begin{align}
    \label{eq:joint_pmf}
    \begin{split}
    &p_\mu(x^T, y^T, u^{T-1})
    = \rho(x_0) B^{x_0,y_0} \\
    &\qquad\qquad\times \prod_{k = 0}^{T-1} A^{x_{k+1},x_k}(u_k) \mu_k^{i_k}(u_k) B^{x_{k+1}, y_{k+1}}(u_k),
    \end{split}
\end{align}
for $T > 0$ where $\prod_{k = 0}^{-1}$ is taken as the identity matrix.
We denote expectation under $p_\mu$ as $E_\mu [\cdot]$.
A policy $\mu = \{ \mu_k^{i_k} : k \geq 0\}$ is \emph{deterministic} if, at all times $k \geq 0$, the support of $\mu_k^{i_k}$ is concentrated at a single control $u_k$; otherwise $\mu$ is \emph{stochastic}.
Let the set of all policies (stochastic or deterministic) be $\mathcal{P}$.

To introduce our ERPOMDP problem, let us define the smoother entropy for $T \geq 0$ under a policy $\mu \in \mathcal{P}$ as
\begin{align}   
    \label{eq:smootherEntropy}
    H_\mu(X^T | Y^T, U^{T-1})
    &\triangleq -E_\mu [ \log p_\mu (X^T | Y^T, U^{T-1})],
\end{align}
and let us define the input-output entropy under $\mu$ as
\begin{align}
    \label{eq:ioEntropy}
    H_\mu(Y^T, U^{T-1})
    &\triangleq -E_\mu [ \log p_\mu (Y^T, U^{T-1})].
\end{align}
Let us also define the additive cost functional
\begin{align}
    \label{eq:costFunctional}
    J_{\mu}^T
    &\triangleq E_\mu \left[ c_T(X_T) + \sum_{k = 0}^{T-1} c \left(X_k, U_k \right) \right]
\end{align}
where $c : \mathcal{X} \times \mathcal{U} \mapsto \mathbb{R}$ and $c_T : \mathcal{X} \mapsto \mathbb{R}$ are arbitrary cost functions dependent on the state and control values.

Our ERPOMDP problem is to find a policy that solves
\begin{align}
    \label{eq:er}
    \begin{aligned}
    & \inf_{\mu \in \mathcal{P}} & & E_T [ J_{\mu}^T + \beta H_\mu(X^T | Y^T, U^{T-1}) + \lambda H_\mu(Y^T, U^{T-1})]\\
    &\mathrm{s.t.} & &  X_{k+1} | X_k, U_k \sim A^{x_{k+1},x_k}(u_k), \quad X_0 \sim \rho\\
    & & & Y_{k+1} | X_{k+1}, U_k \sim B^{x_{k+1},y_{k+1}}(u_k), \quad Y_0 | X_0 \sim B^{x_0,y_0}\\
    & & & U_k | I_k \sim \mu_k^{i_k}(u_k)
    \end{aligned}
    \end{align}
for given nonnegative constants $\beta,\lambda \geq 0$ where the horizon $T \geq 0$ is a random variable with a geometric distribution with (probability of nonoccurence) parameter $0 < \gamma < 1$ such that $\zeta_t \triangleq P(T = t) = \gamma^{t}(1 - \gamma)$ for $t \geq 0$.
{ Despite the entropies above not being in additive forms, we shall show later that the total cost over a geometrically distributed finite horizon is equivalent to a discounted additive cost over an infinite horizon, with $\gamma$ being the discount factor (cf.\ \cite{Shwartz2001}).
}
%\footnote{The choice of a geometric distribution for $T$ is consistent with the principle of maximum entropy on the positive integers when only the mean value is known (such as the mean-time-to-failure in condition monitoring and fault detection, or the estimated-time-of-arrival in navigation). The geometric distribution also has the intuitively appealing memoryless property, $p(T = k + n | T > n) = p(T = k)$ for all $k,n \geq 1$, i.e.\ the probability of having to wait a further $k$ instants for the horizon to end is unaffected by the time $n$ that has passed so far.}

{
The motivation behind our ERPOMDP problem \eqref{eq:er} is twofold.
Firstly, the input-output entropy has an interpretation as the minimum expected number of bits required to transmit or store the observations and controls $(Y^T, U^{T-1})$ (via Shannon's source coding theorem \cite[Section 5.5]{Cover2006}). 
Solving \eqref{eq:er} with $\lambda > 0$ (and $\beta \geq 0$) thus leads to policies that reduce the number of bits used for feedback control (similar to the MDPs in \cite{Tanaka2021}).
Secondly, the smoother entropy intuitively describes the uncertainty associated with estimates of the states $X^T$ given the observations and controls.
Solving \eqref{eq:er} with $\beta > 0$ and $\lambda > 0$ thus leads to policies that reduce the number of bits used to store the observations and controls, whilst ensuring that the states can still be estimated from them.}
%Firstly, the input-output entropy intuitively describes the ``predictablity'' of future observations and controls given past observations and controls (which is important when predictions about the operation of the system must be made such as in system identification or hypothesis testing).
%Similarly, the smoother entropy intuitively describes the uncertainty associated with the states given the observations and controls.
%Solving \eqref{eq:er} with $\beta, \lambda > 0$ thus encourages the selection of policies that jointly reduce uncertainty associated with the states, observations, and controls to make them easier to infer and/or predict (with similar entropy combinations previously considered for the degenerate constant-state POMDP problem of Bayesian experimental design, cf.\ \cite{Chaloner1995,Ryan2016}).
%Secondly, the input-output entropy has an interpretation as the minimum expected number of bits required to represent the observations and controls $(Y^T, U^{T-1})$. 
%Solving \eqref{eq:er} with $\lambda > 0$ (and $\beta \geq 0$) thus encourages policies to reduce the number of bits used for feedback control (similar to the MDPs in \cite{Tanaka2021}).
Operational interpretations of \eqref{eq:er} are discussed further in Section \ref{sec:interp}.

Solving \eqref{eq:er} is greatly simplified if we are able to use standard POMDP solution techniques since they are increasingly able to handle large-scale problems (cf.\ \cite{Kurniawati2008,Araya2010,Garg2019,Fehr2018}).
As discussed in \cite{Araya2010} and \cite[Chapter 8]{Krishnamurthy2016}, the use of standard POMDP techniques to find bounded-error solutions to \eqref{eq:er} requires that: 1) its cost function can be written as an additive function of a sufficient statistic of the information state known as the \emph{belief state}; 2) it can be reformulated as a (fully observed) MDP in terms of the belief state with cost functions that can be arbitrarily well-approximated by PWLC functions; and 3) it can be solved by deterministic policies.
In \cite{Molloy2021a,Molloy2021b}, we showed that this solution approach is possible without the input-output entropy (i.e.\ when $\beta > 0 = \lambda$) by establishing a belief-state expression of the smoother entropy.
The input-output entropy appears more challenging to optimize since its naive factorization as $H_\mu(Y^T, U^{T-1}) =  H_\mu(Y^T | U^{T-1}) + H_\mu(U^{T-1})$ shows immediately that it involves the (unconditional) entropy of the policies $H_\mu(U^{T-1})$, which means that we must consider the possibility of optimal policies solving \eqref{eq:er} being stochastic.
We shall therefore focus on: 1) establishing a belief-state expression of the input-output entropy; 2) showing that it suffices to consider deterministic policies in solving \eqref{eq:er}; and, 3) developing belief MDP reformulations of \eqref{eq:er}.

\section{Belief-State Forms and MDP Reformulation}
\label{sec:reforms}

In this section, we revisit the concept of the belief state and a belief-state form of the smoother entropy.
We then establish a novel belief-state form of the input-output entropy that enables \eqref{eq:er} to be reformulated as a belief MDP amenable to bounded-error solution using standard POMDP techniques.

\subsection{Belief State and Smoother Entropy}

Let $\pi_{k} \in \Delta$ with $\pi_{k}(x) \triangleq p(X_{k} = x | y^k, u^{k-1})$ for $x \in \mathcal{X}$ be the belief state, which evolves via the Bayesian filter:
\begin{align*}
    \pi_{k+1}(x)
    &= \dfrac{ B^{x,y_{k+1}}(u_{k}) \sum_{\bar{x} \in \mathcal{X}} A^{x, \bar{x}}(u_{k}) \pi_{k}(\bar{x})}{ { \sum_{\underline{x} \in \mathcal{X}} \sum_{\bar{x} \in \mathcal{X}} B^{\underline{x},y_{k+1}} (u_{k}) A^{\underline{x}, \bar{x}}(u_{k}) \pi_{k}(\bar{x})}}
\end{align*}
for $k \geq 0$ with $\pi_0(x) = B^{x,y_0}\rho(x) / (\sum_{\bar{x} \in \mathcal{X}} B^{\bar{x},y_0}\rho(\bar{x}))$ for $x \in \mathcal{X}$.
We write the filter as $\pi_{k+1} = \Pi(\pi_{k}, u_{k}, y_{k+1})$.

In \cite{Molloy2021a, Molloy2021b}, we showed that the smoother entropy satisfies
\begin{align}\label{eq:smoother_ent_belief}
 H_\mu(X^T | Y^T, U^{T-1})
 = E_\mu \left[ \tilde{G}_1(\pi_T) + \sum_{k = 0}^{T-1} \tilde{G}_2 (\pi_{k}, U_{k}) \right]
\end{align} {
where $\tilde{G}_1 (\pi_k) \triangleq - \sum_{x \in \mathcal{X}} \pi_k(x) \log \pi_k(x)$ is the \emph{belief-state entropy}, i.e.\ $H_\mu(X_k | y^k, u^{k-1})$, and
\begin{align*}
    \tilde{G}_2(\pi_k, u_k)
    &\triangleq \sum_{x,\bar{x} \in \mathcal{X}} A^{x,\bar{x}}(u_{k}) \pi_{k}(\bar{x}) \log \sum_{\underline{x} \in \mathcal{X}} \dfrac{A^{x,\underline{x}}(u_{k}) \pi_{k}(\underline{x})}{A^{x,\bar{x}}(u_{k}) \pi_{k}(\bar{x})}\\
    &= H_\mu(X_{k}, X_{k+1} | y^{k}, u^{k}) - H_\mu(X_{k+1} | y^{k}, u^{k})
\end{align*}
is the difference between the entropy of $p_\mu(x_k, x_{k+1} | y^k, u^k) = A^{x_{k+1}, x_k}(u_{k}) \pi_k(x_k)$, i.e.\ $H_\mu(X_{k}, X_{k+1} | y^{k}, u^{k})$, and the entropy of $p_\mu(x_{k+1} | y^k, u^k) = \sum_{x_k \in \mathcal{X}} A^{x_{k+1}, x_k}(u_{k}) \pi_k(x_k)$, i.e.\ $H_\mu(X_{k+1} | y^{k}, u^{k})$, with these pmfs computed in the prediction step of the Bayesian filter.
Specifically, the belief-state expression \eqref{eq:smoother_ent_belief} arises because the pmf $p_\mu(x^T | y^T, u^{T-1})$ in \eqref{eq:smootherEntropy} factorizes as
\begin{align}
    \label{eq:backwardFactorisation}
    p_\mu(x^T | y^T, u^{T-1})
    &= \prod_{k = 0}^{T} p_\mu(x_k | x_{k+1}^T, y^{T}, u^{T-1})
\end{align}
via the chain rule with $x_{k+1}^T \triangleq \{x_{k+1}, \ldots, x_T \}$, $x_{T+1}^T \triangleq \emptyset$, and since $p_\mu(x_k | x_{k+1}^T, y^{T}, u^{T-1}) = p_\mu(x_k | x_{k+1}, y^{k}, u^{k}) = p_\mu(x_k, x_{k+1} | y^k, u^k)/p_\mu(x_{k+1} | y^k, u^k)$ via the Markov property of the state and the structure of the measurement kernel and control policy.}
To reformulate \eqref{eq:er} as a belief MDP, we need a similar expression for the input-output entropy.

\subsection{Belief-State Form of Input-Output Entropy}
To establish a novel belief-state form of the input-output entropy \eqref{eq:ioEntropy}, we employ \emph{causally conditioned entropies} as introduced by Kramer \cite{Kramer1998}.
Let the causally conditioned entropy of $Y^T$ given $U^{T-1}$ under any policy $\mu \in \mathcal{P}$ be
\begin{align}
    \label{eq:causalObservationEntropy}
    H_\mu(Y^T \| U^{T-1})
    &\triangleq \sum_{k = 0}^T H_\mu(Y_k | Y^{k-1}, U^{k-1})
\end{align}
where $H_\mu(Y_0 | Y^{-1}, U^{-1}) \triangleq H(Y_0)$ is independent of $\mu$.
Similarly, let the causally conditioned entropy of $U^{T-1}$ given $Y^{T-1}$ under any policy $\mu \in \mathcal{P}$ be
\begin{align}
    \label{eq:causalControlEntropy}
    H_\mu(U^{T-1} \| Y^{T-1})
    &\triangleq \sum_{k = 0}^{T-1} H_\mu(U_k | U^{k-1}, Y^k)
    %   &=  - \sum_{k = 0}^{T-1} E_\mu[ \log \mu_k^{I_k}(U_k)]
\end{align}
with $H_\mu(U_0 | U^{-1}, Y^0) \triangleq H_\mu(U_0 | Y_0)$, $H_\mu(U^{-1} \| Y^{-1}) \triangleq 0$.
Intuitively, $H_\mu(Y^T \| U^{T-1})$ describes the uncertainty associated with the observations given the information causally gained from the controls, whilst $H_\mu(U^{T-1} \| Y^{T-1})$ describes the uncertainty associated with the controls given the information causally gained from the (past) observations.
The following lemma shows that the input-output entropy is the sum of these two causally conditioned entropies.

\begin{lemma}
 \label{lemma:causal_entropy}
 For any $\mu \in \mathcal{P}$ and $T \geq 0$, we have that:
  \begin{align*}
    H_\mu(Y^T, U^{T-1})
    &= H_\mu(Y^T \| U^{T-1}) + H_\mu(U^{T-1} \| Y^{T-1}).
 \end{align*}
\end{lemma}
\begin{IEEEproof}
The proof is via induction.
Note first that $
    H_\mu(Y^0 \| U^{-1}) + H_\mu(U^{-1} \| Y^{-1})
    = H(Y_0)$ proving the lemma assertion for $T = 0$.
Assuming that the lemma assertion holds for trajectories shorter than some length $T > 0$, we now consider it for $T$.
From \eqref{eq:causalObservationEntropy} and \eqref{eq:causalControlEntropy},
\begin{align*}
    &H_\mu(Y^T \| U^{T-1}) + H_\mu(U^{T-1} \| Y^{T-1})\\
    &= H_\mu(Y^{T-1} \| U^{T-2}) + H_\mu(U^{T-2} \| Y^{T-2}) \\
    &\quad+ H_\mu(Y_T | Y^{T-1}, U^{T-1}) + H_\mu(U_{T-1} | U^{T-2}, Y^{T-1})\\
    &= H_\mu(Y^{T-1}, U^{T-2}) + H_\mu(Y_T | Y^{T-1}, U^{T-1}) \\
    &\quad+ H_\mu(U_{T-1} | U^{T-2}, Y^{T-1})
    %&= H(Y^{T-1}, U^{T-1}) + H(Y_T | Y^{T-1}, U^{T-1})\\
    = H_\mu(Y^T, U^{T-1})
\end{align*}
where the second equality holds via the induction hypothesis, and the last equality holds due to the chain rule for conditional entropy.
The proof of the lemma via induction is complete.
\end{IEEEproof}

Lemma \ref{lemma:causal_entropy} differs from trivial expressions of the input-output entropy such as $H_\mu(Y^T, U^{T-1}) = H_\mu(Y^T) + H_\mu(U^{T-1} | Y^T) = H_\mu(U^{T-1}) + H_\mu(Y^T | U^{T-1})$ since these involve conditional entropies conditioned on the entire trajectories $Y^T$ and $U^{T-1}$, whilst Lemma \ref{lemma:causal_entropy} establishes a form involving sums of conditional entropies only conditioned on the information state $I_k$ at each time $k$.
Lemma \ref{lemma:causal_entropy} thus leads to a belief-state expression of the input-output entropy.
Specifically, the definition of $H_\mu(Y^T \| U^{T-1})$ in \eqref{eq:causalObservationEntropy} and the tower property of conditional expectation gives that
\begin{align}\label{eq:io_ent_belief}
    H_\mu(Y^T \| U^{T-1})
    &= H(Y_0) + E_\mu \left[ \sum_{k = 0}^{T-1} \tilde{G}_3(\pi_k, U_k) \right]
\end{align}
where $\tilde{G}_3(\pi_k, u_k)$ is the entropy of the conditional pmf 
\begin{align}
    \label{eq:beliefObs}
    p(y_{k+1} | \pi_k, u_k) 
    &= \sum_{x, \bar{x} \in \mathcal{X}} B^{x,y_{k+1}} (u_{k}) A^{x, \bar{x}}(u_{k}) \pi_{k}(\bar{x}),
\end{align}
that is, $H (Y_{k+1} | y^k, u^k)$, defined as
\begin{align}
    \label{eq:g3}
    \tilde{G}_3(\pi_k, u_k)
    \triangleq - \sum_{y \in \mathcal{Y}} p(y | \pi_k, u_k) \log p(y | \pi_k, u_k).
\end{align}
A similar belief-state form of $H_\mu(U^{T-1} \| Y^{T-1})$ also holds but will prove unnecessary since we shall next show that deterministic policies solve \eqref{eq:er} (for which $H_\mu(U^{T-1} \| Y^{T-1}) = 0$). % so that $H_\mu(Y^T, U^{T-1}) = H_\mu(Y^T \| U^{T-1})$).

\subsection{Belief MDP Reformulation}

Along the lines of considering deterministic policies and omitting $H_\mu(U^{T-1} \| Y^{T-1})$, the following lemma introduces a useful surrogate problem and is the final intermediate result we require to reformulate \eqref{eq:er} as a belief MDP.

\begin{lemma}
\label{lemma:simplified_er}
    If a deterministic policy $\mu^* \in \mathcal{P}$ minimizes
\begin{align}
\label{eq:simplified_er}
 \begin{split}
    E_T [ J_{\mu}^T + \beta H_\mu(X^T | Y^T, U^{T-1}) + \lambda H_{\mu}(Y^T \| U^{T-1})]
 \end{split}
\end{align}
over all policies $\mu \in \mathcal{P}$ under the same constraints as \eqref{eq:er} given $\beta,\lambda \geq 0$, then $\mu^*$ also solves \eqref{eq:er} with the same $\beta,\lambda \geq 0$.
\end{lemma}
\begin{IEEEproof}
    The definition of the infimum implies that
\begin{align*}
    &E_T[J_{\mu^*}^T + \beta H_{\mu^*}(X^T | Y^T, U^{T-1}) + \lambda H_{\mu^*}(Y^T, U^{T-1})]\\
    &\geq \inf_{\mu \in \mathcal{P}} E_T[ J_{\mu}^T + \beta H_{\mu}(X^T | Y^T, U^{T-1}) + \lambda H_\mu(Y^T, U^{T-1})]\\
    &\geq \inf_{\mu \in \mathcal{P}} E_T[J_{\mu}^T + \beta H_\mu(X^T | Y^T, U^{T-1}) + \lambda H_\mu(Y^T \| U^{T-1})]\\
    %&\geq \inf_{\mu \in \mathcal{P}} \left\{J_{\mu}^T + \lambda H_\mu(Y^T \| U^{T-1}) \right\}\\
    &= E_T [ J_{\mu^*}^T + \beta H_{\mu^*}(X^T | Y^T, U^{T-1}) + \lambda H_{\mu^*}(Y^T \| U^{T-1})]
\end{align*}
where the second inequality holds due to Lemma \ref{lemma:causal_entropy} by noting that $H_{\mu}(U^{T-1} \| Y^{T-1}) \geq 0$ for all $\mu \in \mathcal{P}$, and the last line holds via the definition of $\mu^*$.
These inequalities must hold with equality since Lemma \ref{lemma:causal_entropy} combined with $H_{\mu^*}(U^{T-1} \| Y^{T-1}) = 0$ due to $\mu^*$ being deterministic implies
\begin{align*}
    &E_T[J_{\mu^*}^T + \beta H_{\mu^*} (X^T | Y^T, U^{T-1}) + \lambda H_{\mu^*}(Y^T, U^{T-1})]\\
    %&= J_{\mu^*}^T + \lambda H_{\mu^*}(Y^T \| U^{T-1}) + \lambda H_{\mu^*}(U^{T-1} \| Y^{T-1})\\
    &= E_T[J_{\mu^*}^T + \beta H_{\mu^*} (X^T | Y^T, U^{T-1}) + \lambda H_{\mu^*}(Y^T \| U^{T-1})].
\end{align*}
The proof is complete.
\end{IEEEproof}

A reformulation of \eqref{eq:er} as a belief MDP follows.

\begin{theorem}
\label{theorem:MDP}
Define the belief-state cost function
\begin{align}
    \label{eq:belief_cost}
    \begin{split}
    &G(\pi_k, u_k)\\
    &\triangleq (1 - \gamma) \beta \tilde{G}_1(\pi_k) + \gamma \beta \tilde{G}_2(\pi_k, u_k) + \gamma \lambda \tilde{G}_3(\pi_k, u_k)\\
    &\quad + E_{X_k} [(1 - \gamma)c_T(X_k) + \gamma c(X_k, U_k) | \pi_k, U_k = u_k].
    \end{split}
\end{align}
Then \eqref{eq:er} with $\beta,\lambda \geq 0$ is equivalent (up to $\lambda H(Y_0)$) to:
\begin{align}
\label{eq:er_mdp}
\begin{aligned}
&\inf_{\bar{\mu}} & & E_{\bar{\mu}} \left[ \left. \sum_{k = 0}^{\infty} \gamma^k G \left( \pi_{k}, U_{k} \right) \right| \pi_0 \right]\\ 
&\mathrm{s.t.} & &  \pi_{k+1} = \Pi\left( \pi_{k}, U_{k}, Y_{k+1} \right)\\
& & & Y_{k+1} | \pi_k, U_k \sim p(y_{k+1} | \pi_k, u_k)\\
& & & U_k = \bar{\mu}(\pi_k) \in \mathcal{U}
\end{aligned}
\end{align}
where the optimization is over deterministic, stationary policies $\bar{\mu} : \Delta \mapsto \mathcal{U}$ that are functions of the belief state $\pi_k$, and { $\gamma$ is the parameter of the geometric distribution of $T$.}
\end{theorem}
\begin{IEEEproof}
Given Lemma \ref{lemma:simplified_er}, it suffices to show that minimizing \eqref{eq:simplified_er} under the same constraints as \eqref{eq:er} is equivalent (up to the constant $\lambda H(Y_0)$) to the belief MDP \eqref{eq:er_mdp}.

Rewriting \eqref{eq:simplified_er} for any $\mu \in \mathcal{P}$ using \eqref{eq:costFunctional}, \eqref{eq:smoother_ent_belief}, and \eqref{eq:io_ent_belief} gives
\begin{align*}
    &E_T [ J_{\mu}^T + \beta H_\mu(X^T | Y^T, U^{T-1}) + \lambda H_\mu(Y^T \| U^{T-1})]\\
    %&= \lambda H(Y_0) + E_{T,\mu} \Bigg[ c_T(X_T) + \beta \tilde{G}_1(\pi_T) \\
    %&\quad+ \sum_{k = 0}^{T-1} \left[ c(X_k, U_k) + \beta \tilde{G}_2(\pi_k, U_k) + \lambda \tilde{G}_3(X_k, U_k) \right] \Bigg]\\
    &= \lambda H(Y_0) + E_{T,\mu} \Bigg[ \tilde{G}_T(\pi_T) + \sum_{k = 0}^{T-1} \tilde{G}(\pi_k, U_k) \Bigg]
\end{align*}
via nested expectations with $\tilde{G}_T(\pi_T) \triangleq E_{X_T} [ c_T(X_T) + \beta \tilde{G}_1(\pi_T) | \pi_T ]$ and $\tilde{G}(\pi_k, U_k) \triangleq E_{X_k} [ c(X_k, U_k) + \beta \tilde{G}_2(\pi_k, U_k) + \lambda \tilde{G}_3(\pi_k, U_k) | \pi_k, U_k ]$.
Ignoring $\lambda H(Y_0)$,
\begin{align*}
    &E_{T,\mu} \Bigg[ \tilde{G}_T(\pi_T) + \sum_{k = 0}^{T-1} \tilde{G}(\pi_k, U_k) \Bigg]\\
    %&= E_\mu \Bigg[ E_T \Bigg[ \tilde{G}_T(\pi_T) + \sum_{k = 0}^{T-1} \tilde{G}(\pi_k, U_k) \Bigg] \Bigg]\\
    &= E_\mu \left[ \sum_{t = 0}^\infty \zeta_t \left( \tilde{G}_T(\pi_t) + \sum_{k = 0}^{t-1} \tilde{G}\left(\pi_k, U_k \right) \right) \right] \\
    &= E_\mu \left[ \sum_{k = 0}^\infty \left( \zeta_k \tilde{G}_T(\pi_k)  + \sum_{t = k+1}^\infty  \zeta_t \tilde{G} \left(\pi_k, U_k \right) \right) \right] \\
    &= E_\mu \left[ \sum_{k = 0}^\infty ( \zeta_k \tilde{G}_T(\pi_k)  + P(T > k) \tilde{G} \left(\pi_k, U_k \right) )  \right]\\
    &= E_\mu \left[ \sum_{k = 0}^\infty \gamma^k \left( (1 - \gamma) \tilde{G}_T(\pi_k) + \gamma \tilde{G}(\pi_k, U_k) \right) \right]\\
    &= E_\mu \left[ \sum_{k = 0}^\infty \gamma^k G(\pi_k, U_k) \right]
\end{align*}
where the second equality holds by interchanging summations; the third and fourth equalities follow from the cumulative distribution and pmf of the geometric distribution; and, the last equality holds by definition.
Standard POMDP (or MDP) results imply that this expectation can be minimized over $\mu \in \mathcal{P}$ under the same constraints as \eqref{eq:er} by deterministic stationary policies $\bar{\mu}$ that are functions of $\pi_k$  (cf.\ \cite[Section 5.4.1]{Bertsekas2005} and \cite[Theorem 6.2.2]{Krishnamurthy2016}). The proof is complete.
\end{IEEEproof}

\subsection{Structural Results and Bounded-Error Solutions}

Given the reformulation of \eqref{eq:er} in Theorem \ref{theorem:MDP}, standard MDP or POMDP results (e.g., \cite[Theorem 6.2.2]{Krishnamurthy2016} or \cite{Araya2010}) imply that an optimal policy $\bar{\mu}^* : \Delta \mapsto \mathcal{U}$ and value function $V : \Delta \mapsto \mathbb{R}$ solving \eqref{eq:er} satisfy Bellman's equation
\begin{align}
    \label{eq:bellman}
    V(\pi)
    = \min_{u \in \mathcal{U}} \left\{ G(\pi, u) + \gamma E_{Y} \left[ V(\Pi(\pi, u, Y)) | \pi, u \right] \right\}
\end{align}
for all $\pi \in \Delta$ with $\bar{\mu}^*(\pi)$ being a minimizing argument of \eqref{eq:bellman}.
%\begin{align*}
%    &\bar{\mu}^*(\pi_k)\\
%    &\in \argmin_{u \in \mathcal{U}} \left\{ G(\pi_k, u) + \gamma E_{Y_{k+1}} \left[ V(\Pi(\pi_k, u, Y_{k+1})) | \pi_k, u \right] \right\}.
%\end{align*}
Solving \eqref{eq:bellman} is, in general, difficult.
However, if the functions $G$ and $V$ are concave in $\pi$, then standard POMDP techniques can yield solutions to \eqref{eq:bellman}.
We thus examine $G$ and $V$.

\begin{theorem}
\label{theorem:concave}
The cost and value functions $G( \pi_k, u_k)$ and $V(\pi_k)$ of \eqref{eq:er} reformulated as \eqref{eq:er_mdp} are concave and continuous in $\pi_k \in \Delta$ for all $u_k \in \mathcal{U}$, all $0 < \gamma < 1$, and all $\beta,\lambda \geq 0$.
\end{theorem}
\begin{IEEEproof}
    To prove the theorem assertion for $G$, it suffices to show that each function in \eqref{eq:belief_cost} is concave and continuous in $\pi_k$ since their coefficients are nonnegative for $\beta,\lambda \geq 0$ and $0 < \gamma < 1$.
	Firstly, $\tilde{G}_1$ and $\tilde{G}_2$ are concave and continuous in $\pi_k$ via \cite[Lemma 2]{Molloy2021a}.
	Secondly, $\tilde{G}_3$ is the entropy of the conditional pmf $p(y | \pi_k, u_k)$, and so is concave and continuous in it via \cite[Theorem 2.7.3]{Cover2006}.
	Since $p(y | \pi_k, u_k)$ is linear in $\pi_k$ (cf.\ \eqref{eq:beliefObs}), $\tilde{G}_3$ is concave in a linear function of $\pi_k$, and so is concave and continuous in $\pi_k$.
	Finally, the expectation in \eqref{eq:belief_cost} is concave and continuous in $\pi_k$ since it equals
	$
     \sum_{x \in \mathcal{X}} \pi_k(x) \left[ (1 - \gamma)c_T(x) + \gamma c(x, u_k) \right].
	$
	That $V$ is concave and continuous follows via \cite[Theorem 3.1]{Araya2010}.
\end{IEEEproof}

Theorem \ref{theorem:concave} enables the use of standard POMDP techniques to find bounded-error solutions to \eqref{eq:er}.
Specifically, following the PWLC approach proposed in \cite{Araya2010}, consider an arbitrary finite set $\Xi \subset \Delta$ of \emph{base points} $\xi \in \Xi$ at which the gradient $\nabla_\pi G(\xi, u)$ of $G(\cdot, u)$ is well defined for all $u \in \mathcal{U}$.
For each $u \in \mathcal{U}$, the tangent hyperplane to $G(\cdot, u)$ at each $\xi \in \Xi$ is
\begin{align*}
    \omega_\xi^u (\pi)
    \triangleq G(\xi, u) + \left< (\pi - \xi), \nabla_\pi G(\xi, u) \right>
    = \left< \pi, \alpha_\xi^u \right>
\end{align*}
for $\pi \in \Delta$ where $\left< \cdot, \cdot \right>$ denotes the inner product, and $\alpha_\xi^u \triangleq G(\xi, u) + \nabla_\pi G(\xi, u) - \left< \xi, \nabla_\pi G(\xi, u) \right> \in \mathbb{R}^{N_x}$ are vectors (with the addition of a vector and a scalar here meaning the addition of the scalar to all components of the vector).
Since $G$ is concave via Theorem \ref{theorem:concave}, the hyperplanes form a PWLC approximation $\hat{G}$ to $G$; that is, for $\pi \in \Delta$ and $u \in \mathcal{U}$,
\begin{align*}
    \hat{G}(\pi, u) \triangleq \min_{\xi \in \Xi} \left< \pi, \alpha_\xi^u \right> \geq G(\pi, u).
\end{align*}

By replacing $G$ in \eqref{eq:bellman} with $\hat{G}$, \eqref{eq:bellman} can be solved for an approximate PWLC value function $\hat{V}$ (and policy) using standard POMDP algorithms that operate directly on the vectors $\{\alpha_{\xi}^{u} : \xi \in \Xi,\, u \in \mathcal{U}\}$ (see \cite[Section 3.3]{Araya2010} for more details).
{ Furthermore, $G$ satisfies the H{\"o}lder continuity condition of \cite[Theorem 4.3]{Araya2010} since the (negative) entropy function $f(x) = \sum_{i = 1}^{N_x} x(i) \log x(i)$ is H{\"o}lder continuous on $\Delta$ (as are continuous linear functions, and the sums and compositions of H{\"o}lder continuous functions,  cf. \cite[Example 1.1.4, and Propositions 1.2.1 and 1.2.2]{Fiorenza2017}).
Hence, \cite[Section 4.2]{Araya2010}} implies that there exists constants $\kappa_1 > 0$ and $\kappa_2 \in (0,1)$ such that
$
    \| V - \hat{V} \|_\infty
    \leq \kappa_1 (\delta_\Xi)^{\kappa_2}
$
where $\delta_\Xi \triangleq \min_{\pi \in \Delta} \max_{\xi \in \Xi} \| \pi - \xi \|_1$ is the sparsity of $\Xi$ with $\|\cdot\|_1$ and $\|\cdot\|_\infty$ denoting the $l^1$-norm and $L^\infty$-norm, respectively.
In principle, this error can be made arbitrarily small by selecting $\xi \in \Xi$ to decrease $\delta_\Xi$.

% Then there exists scalar constants $\kappa_1 > 0$ and $\kappa_2 \in (0,1)$ such that the errors in the PWLC approximations are bounded in the sense that
%     $
%         |G(\pi,u) - \hat{G}(\pi,u)| \leq \kappa_1(\delta_\Xi)^{\kappa_2}
%     $
% for all $\pi \in \Delta^N$ and all $u \in \mathcal{U}$ where $\delta_\Xi \triangleq \min_{\pi \in \Delta} \max_{\xi \in \Xi} \| \pi - \xi \|_1$ is the sparsity of the base-point set $\Xi$ and $\|\cdot\|_1$ denotes the $l^1$-norm.

\section{Special Case of Joint-Entropy Regularization}
\label{sec:jer}

The PWLC approach to solving \eqref{eq:er} presented in the previous section is consistent with state-of-the-art approaches to solving POMDPs with nonlinear belief-state cost functions (see \cite[Chapter 8]{Krishnamurthy2016}, \cite{Fehr2018,Araya2010}).
However, constructing accurate PWLC approximations can require a large number of linear segments (i.e.\ vectors $\alpha_\xi^u$), resulting in significant computational effort and the need to modify standard POMDP solver implementations (cf.\ \cite[Section 8.4.5]{Krishnamurthy2016}). % and \cite{Dressel2017}).
In this section, we explore a simpler approach to solving \eqref{eq:er} that is tractable \emph{without} PWLC approximations when the smoother and input-output entropies are equally penalized, that is, when $\beta = \lambda \geq 0$ so that the sum of the smoother and input-output entropies in \eqref{eq:er} becomes the \emph{joint entropy} defined as
$
    H_\mu(X^T, Y^T, U^{T-1})
    \triangleq - E_\mu[ \log p_\mu ( X^T, Y^T, U^{T-1}) ]
    = H_\mu(X^T | Y^T, U^{T-1}) + H_\mu(Y^T, U^{T-1})$.
Key to this simpler approach is the following new expression for the joint entropy.

\begin{lemma}
 \label{lemma:joint_entropy_stochastic}
 For any policy $\mu \in \mathcal{P}$ and $T \geq 0$, we have:
  \begin{align*}
    &H_\mu(X^T, Y^T, U^{T-1})\\
    &= H(X_0, Y_0) + H_\mu (U^{T-1} \| Y^{T-1})
    + E_\mu \Bigg[ \sum_{k = 0}^{T-1} \tilde{c}(X_k, U_k) \Bigg] 
 \end{align*}
 where 
 \begin{align*}
    \begin{split}
     &\tilde{c} (x_k, u_k)\\
     &\triangleq - \sum_{x \in \mathcal{X}} \sum_{y \in \mathcal{Y}} A^{x, x_k}(u_k) B^{x,y} (u_k) \log (A^{x, x_k}(u_k) B^{x,y}(u_k)).
    \end{split}
 \end{align*}
\end{lemma}
\begin{IEEEproof}
From the definition of the joint entropy and \eqref{eq:joint_pmf},
  \begin{align*}
      &H_\mu(X^T, Y^T, U^{T-1})\\
      %&= - E_\mu \left[ \log p_\mu(X^T, Y^T, U^{T-1}) \right]\\
      &= - E_\mu \Bigg[ \log (\rho(X_0) B^{X_0,Y_0})
      + \sum_{k = 0}^{T-1} \log (\mu_k^{I_k}(U_k) Z_{k}) \Bigg]\\
      &= H(X_0,Y_0) - \sum_{k = 0}^{T-1} E_\mu \left[ \log \mu_k^{I_k}(U_k) \right]
 - \sum_{k = 0}^{T-1} E_\mu \left[ \log Z_{k} \right]\\
      &= H(X_0,Y_0) + H_\mu(U^{T-1} \| Y^{T-1}) + E_\mu \Bigg[ \sum_{k = 0}^{T-1} \tilde{c}(X_k, U_k) \Bigg]
  \end{align*}
  where $Z_{k} \triangleq A^{X_{k+1},X_k}(U_k) B^{X_{k+1}, Y_{k+1}}(U_k)$; the second equality holds due to the properties of the logarithm and linearity of expectations and summations; and, the third equality follows from \eqref{eq:causalControlEntropy}, with nested expectations giving
  \begin{align*}
    - E_\mu \left[ \log Z_{k} \right]
    = E_\mu \left[ E_{X_{k+1}, Y_{k+1}} \left[ - \left. \log Z_{k}
\right| X_k, U_k \right] \right]
  \end{align*}  
  where the inner expectation is $\tilde{c}$. The proof is complete.
\end{IEEEproof}

A second reformulation of \eqref{eq:er} with $\beta = \lambda \geq 0$ follows.

\begin{theorem}
\label{theorem:jer}
% Define the cost function
% \begin{align*}
%     \ell(x_k, u_k)
%     &\triangleq (1 - \gamma)c_T(x_k) + \gamma c(x_k, u_k) + \gamma \beta \tilde{c}(x_k, u_k) 
% \end{align*}
% then solving \eqref{eq:er} is equivalent to solving the standard POMDP:
%     \begin{align}
%     \label{eq:jem_pomdp}
%     \begin{aligned}
%     &\inf_{\mu} & & E_{\mu} \left[ \sum_{k = 0}^{\infty} \gamma^k \ell \left( X_k, U_k \right) \right]\\ 
%     &\mathrm{s.t.} & &  X_{k+1} | X_k, U_k \sim A^{x_{k+1},x_k}(u_k), X_0 \sim \rho\\
%     & & & Y_{k+1} | X_{k+1}, U_k \sim B^{x_{k+1},y_{k+1}}(u_k)\\
%     & & & U_k = \mu_k^{i_k} \in \mathcal{U}
%     \end{aligned}
% \end{align}
% when $\beta = \lambda \geq 0$ with the optimization over deterministic policies of the information state $I_k$.
Define the belief-state cost function
\begin{align*}
    L(\pi_k,u_k)
    \triangleq E_{X_k} [ \ell(X_k, u_k) | \pi_k, u_k ]
    = \sum_{x \in \mathcal{X}} \pi_k(x) \ell(x,u_k)
\end{align*}
where $\ell(x_k, u_k) \triangleq (1 - \gamma)c_T(x_k) + \gamma c(x_k, u_k) + \gamma \beta \tilde{c}(x_k, u_k)$, then \eqref{eq:er} with $\beta = \lambda \geq 0$ is equivalent (up to $\beta H(X_0,Y_0)$) to:
\begin{align}
\label{eq:jer_mdp}
\begin{aligned}
&\inf_{\bar{\mu}} & & E_{\bar{\mu}} \left[ \left. \sum_{k = 0}^{\infty} \gamma^k L \left( \pi_{k}, U_{k} \right) \right| \pi_0 \right]
%&\mathrm{s.t.} & &  \pi_{k+1} = \Pi\left( \pi_{k}, U_{k}, Y_{k+1} \right)\\
%& & & Y_{k+1} | \pi_k, U_k \sim p(y_{k+1} | \pi_k, u_k)\\
%& & & U_k = \bar{\mu}_k^{\pi_k} \in \mathcal{U}
\end{aligned}
\end{align}
subject to the same constraints as \eqref{eq:er_mdp} and where the optimization is over deterministic, stationary policies $\bar{\mu} : \Delta \mapsto \mathcal{U}$.
\end{theorem}
\begin{IEEEproof}
Same as that of Theorem \ref{theorem:MDP}, but using Lemma \ref{lemma:joint_entropy_stochastic} instead of \eqref{eq:smoother_ent_belief} and \eqref{eq:io_ent_belief} to rewrite \eqref{eq:simplified_er}, noting that $H_\mu(X^T | Y^{T}, U^{T-1}) + H_\mu(Y^T \| U^{T-1}) = H_\mu(X^T, Y^T, U^{T-1}) - H_\mu(U^{T-1} \| Y^{T-1})$ via Lemma \ref{lemma:causal_entropy}.
\end{IEEEproof}

The belief MDP reformulation of our ERPOMDP problem in \eqref{eq:jer_mdp} with $\beta = \lambda$ is surprising because its cost function $L$ is \emph{linear} in $\pi_k$.
In contrast, the cost function $G$ of the first belief MDP reformulation established in \eqref{eq:er_mdp} is \emph{nonlinear} in $\pi_k$, even when $\beta = \lambda$.
The two different belief MDP reformulations of \eqref{eq:er} in \eqref{eq:er_mdp} and \eqref{eq:jer_mdp} are due to the joint pmf $p_\mu(x^T, y^T, u^{T-1})$ admitting multiple factorizations, with \eqref{eq:joint_pmf} leading to \eqref{eq:jer_mdp}, and a factorization similar to \eqref{eq:backwardFactorisation} leading to \eqref{eq:er_mdp}.

%\footnote{We avoid explicit use of this more complicated factorization by exploiting Lemma \ref{lemma:causal_entropy}, however, similar factorizations of (uncontrolled) hidden Markov models can be found in \cite[Eq.\ (1)]{Doucet2000}.} 

The linearity of $L$ is of considerable practical value because it enables \eqref{eq:er} with $\beta = \lambda$ to be solved using standard POMDP solution techniques without any PWLC approximation of $L$.
Indeed, for any $u \in \mathcal{U}$ and $\pi \in \Delta$, $L(\pi, u) = \left<\pi, \alpha^u \right>$ holds exactly given the (single) vector $\alpha^u \triangleq \left[ \ell(1, u), \ell(2, u), \ldots, \ell(N, u) \right]'$.
Dynamic programming equations of the form of \eqref{eq:bellman} with $L$ replacing $G$ can thus be solved using standard POMDP techniques that operate directly on the vectors $\{\alpha^{u} : u \in \mathcal{U}\}$ (cf.\ \cite[Chapter 7.5]{Krishnamurthy2016}).
We next discuss the operational significance of the linearity of $L$.

\section{Operational Interpretations and Relationships}
\label{sec:interp}

In this section, we discuss two operational interpretations of ERPOMDPs, and discuss their relationship to other optimization problems with information-theoretic terms.

\begin{figure}[t!]
     \centering
     \begin{subfigure}[t!]{\columnwidth}
         \centering
         \includegraphics[width = 0.9\columnwidth]{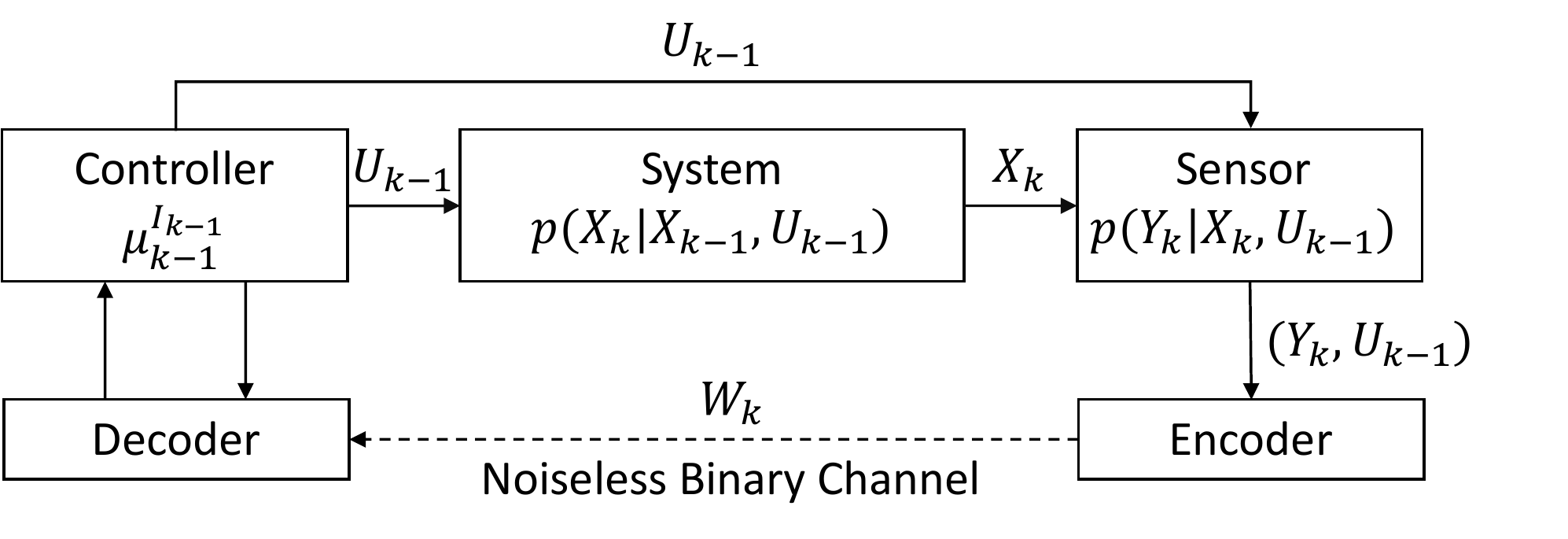}
         \caption{}
         \label{fig:network}
     \end{subfigure}
     \vspace{0.2cm}
     \vfill
     \begin{subfigure}[t!]{\columnwidth}
         \centering
         \includegraphics[width = 0.9\columnwidth]{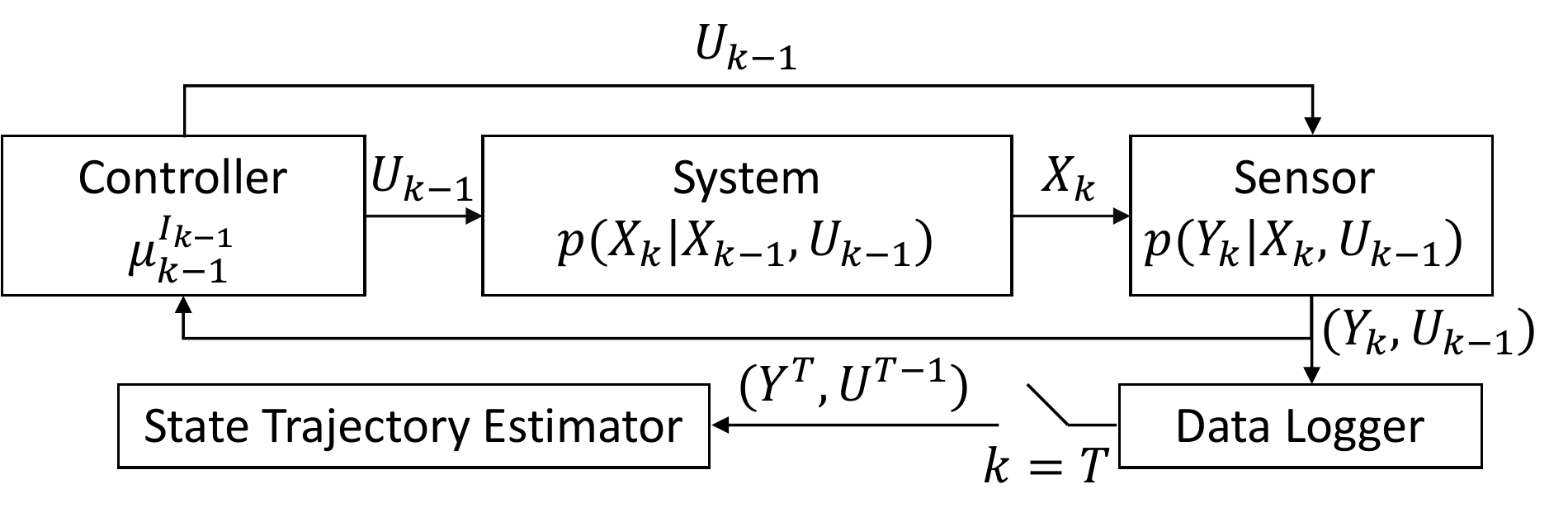}
         \caption{}
         \label{fig:smoothing}
     \end{subfigure}
        \caption{Operational Interpretations of the ERPOMDP problem \eqref{eq:er}: (a) Networked Control; and (b) Memory-Efficient Active State Trajectory Estimation.}
        \label{fig:interps}
        %\vspace{-0.5cm}
\end{figure}

\subsection{Networked Control Rate-Cost Trade-Offs}
Consider a networked control setting in which the feedback path of a POMDP involves transmission over a noiseless binary channel, as illustrated in Fig.\ \ref{fig:network}.
At every time step $k$, an encoder receives observations $Y_k$ of the state $X_k$ (e.g., arising from a sensor), and selects and transmits a binary codeword $W_k$ from a predefined codebook of optimal uniquely decodable binary codes $\mathcal{W}_k$.
Upon receiving $W_k$, the decoder decodes $Y_k$ and passes it to a controller, that uses it to evaluate the next control $U_k$.
We allow the codebook $\mathcal{W}_k$ to be time-varying, and let $R_k$ be the length (number of bits) of the codeword $W_k$.
We assume that both the encoder and decoder have infinite memories so that $W_k$ conveys $Y_k$ given $(Y^{k-1}, U^{k-1})$.
Shannon's source coding theorem (cf.\ \cite[Section 5.5]{Cover2006}) then implies that the (minimum) expected data transmitted in the noiseless binary channel at time $k$ satisfies
\begin{align*}
    H_\mu (Y_k | Y^{k-1}, U^{k-1})
    \leq E_{\mu} \left[ R_k \right]
    \leq H_\mu (Y_k | Y^{k-1}, U^{k-1}) + 1,
\end{align*}
and so the total expected data transmitted over $T$ satisfies
\begin{align*}
    E_T [ H_\mu (Y^T \| U^{T-1})] 
    &\leq E_{T,\mu} \left[ \sum_{k = 0}^T R_k \right] \\
    &\leq E_T \left[ H_\mu (Y^T \| U^{T-1}) + T + 1 \right].
\end{align*}
In view of Lemma \ref{lemma:simplified_er} and Theorem \ref{theorem:MDP}, our ERPOMDP problem \eqref{eq:er} with $\lambda > 0$ (and any $\beta \geq 0$) thus has the operational interpretation of seeking policies that trade-off the expected total data transmitted for feedback control via $H_\mu (Y^T \| U^{T-1})$, with the value of the cost functional $J_\mu^T$.

{ Theorem \ref{theorem:MDP} is particularly important for introducing rate-cost trade-offs into POMDPs because it enables regularization only by the input-output entropy (i.e., it holds when $\lambda > 0$ but $\beta = 0$ in \eqref{eq:er}).
In contrast, Theorem \ref{theorem:jer} is, in general, of secondary value for introducing rate-cost trade-offs because it also requires regularization by the smoother entropy (i.e., it holds only when $\beta = \lambda$ in \eqref{eq:er}), but does enable the use of standard POMDP techniques without PWLC approximations.% may, however, outweigh the drawback of superfluous regularization by the smoother entropy.
}

\subsection{Memory-Efficient Active State Trajectory Estimation}
The second operational interpretation of our ERPOMDP problem \eqref{eq:er} relates to active state estimation (i.e.\ controlling a POMDP to aid the estimation of its latent state trajectory $X^T$ from stored trajectories $(Y^T, U^{T-1})$).
Such problems arise in robotics (cf.\ \cite{Thrun2005, Stachniss2005}) and controlled sensing (cf.\ \cite{Krishnamurthy2007,Krishnamurthy2016}).

As shown in Fig.\ \ref{fig:smoothing}, consider a POMDP in which, at each time $k$, the current observation and control $(Y_k,U_{k-1})$ are encoded and stored by a data logger by selecting and storing a binary codeword $W_k$ from a predefined codebook of optimal uniquely decodable binary codes $\mathcal{W}_k$.
Thus, the data logger encodes $(Y_k,U_{k-1})$ given $(Y^{k-1}, U^{k-2})$.
We allow the codebook $\mathcal{W}_k$ to be time-varying, and let $R_k$ be the length of the codeword $W_k$. 
Shannon's source coding theorem (cf.\ \cite[Section 5.5]{Cover2006}) implies that the (minimum) expected total memory required to store $(Y^T,U^{T-1})$ satisfies the bounds
\begin{align*}
    \begin{split}
    E_T [ H_\mu (Y^T, U^{T-1})] 
    &\leq E_{T,\mu} \left[ \sum_{k = 0}^T R_k \right] \\
    &\leq E_T \left[ H_\mu (Y^T, U^{T-1}) + T + 1 \right].
    \end{split}
\end{align*}

At the conclusion of the control horizon ($k = T$), the data logger decodes the stored trajectories $(Y^T,U^{T-1})$ and passes them to an (offline) algorithm for estimating the state trajectory $X^T$ (e.g.\ the Viterbi algorithm).
Let the minimum probability of error for \emph{any} estimator of $X^T$ given $(Y^T, U^{T-1})$ be
$
	\epsilon
	\triangleq \min_{\hat{X}^T} P(X^T \neq \hat{X}^T)
$
with $\hat{X}^T$ being any function $f : \mathcal{Y}^T \times \mathcal{U}^{T-1} \mapsto \mathcal{X}^T$.
Theorem 1 of \cite{Feder1994} gives that
\begin{align*}
    \Phi^{-1}(H_\mu(X^T | Y^T, U^{T-1}))
    \leq \epsilon
    \leq \phi^{-1}(H_\mu(X^T | Y^T, U^{T-1}))
\end{align*}
where $\Phi^{-1}$ and $\phi^{-1}$ are the inverse functions of strictly monotonically increasing functions (defined in \cite{Feder1994}), and thus are themselves strictly monotonically increasing.
%The smoother entropy has been investigated as a criterion for active state estimation in \cite{Molloy2021a,Stachniss2005}.

The involvement of the smoother and input-output entropies in bounds on the estimation error and required memory implies that \eqref{eq:er} has the operational interpretation of seeking policies that aid estimation of the state trajectory (by reducing the smoother entropy) whilst decreasing the memory required to store the observation and control trajectories (by reducing the input-output entropy).
In the special case considered in Section \ref{sec:jer} with $\beta = \lambda$, \eqref{eq:er} constitutes a formulation of active state estimation in which the estimation and memory objectives are weighted equally.
{ In this regard, Theorem \ref{theorem:jer} establishing the linearity of $L$ in \eqref{eq:jer_mdp} is further surprising since most previous active state estimation formulations involve cost that are entirely nonlinear in the belief state and can only be optimized by resorting to approximations (cf.\ \cite{Krishnamurthy2007,Araya2010,Krishnamurthy2016}).}
%For example, whilst ubiquitous, the entropy of the belief state, $H(X_k | y^k, u^{k-1}) = - \sum_{x \in \mathcal{X}} \pi_k(x) \log \pi_k(x)$, is nonlinear in $\pi_k$, and so its practical use requires approximations (cf.\ \cite{Krishnamurthy2007,Araya2010,Molloy2021a,Krishnamurthy2016}).}

\subsection{Relationship to Other Information-Theoretic POMDPs}
{
ERPOMDPs \eqref{eq:er} are closely related to problems involving the optimization of information-theoretic terms that have previously been considered for reinforcement learning \cite{Haarnoja2017}, studying the capacity of channels with memory and feedback \cite{Charalambous2016}, privacy (e.g. in smart metering systems) \cite{Li2018,Savas2020,Sabag2020,Tanaka2018}, and studying rate-cost trade-offs in MDPs and POMDPs \cite{Tanaka2021,Rubin2012}.
These problems, however, mostly involve optimizing only a single information-theoretic term derived from either the mutual information between states and/or observations (e.g., directed information and transfer entropy \cite{Charalambous2016,Li2018,Tanaka2018,Tanaka2021,Rubin2012,Sabag2020}), or the entropy of the states or controls (e.g., $H_\mu(U_k | i_k) = -\sum_{u \in \mathcal{U}} \mu_k^{i_k}(u) \log \mu_k^{i_k}(u)$ \cite{Haarnoja2017,Savas2020}).
In contrast, ERPOMDPs involve both the standard cost functional $J_\mu^T$ and, in general, two information-theoretic terms, the smoother entropy and the (novel) input-output entropy.
The procedure for solving ERPOMDPs is, however, similar to that of solving these other optimization problems, with most having been shown to lead to belief MDPs --- albeit few (if any) with cost functions that are linear in the belief state, rendering our Theorem \ref{theorem:jer} joint-entropy result further surprising.
}

\section{Simulation Example}
\label{sec:results}

We now simulate ERPOMDPs for active state estimation.

\subsection{Example Set-Up}
\begin{figure}[t!]
     \centering
     \begin{subfigure}[h!]{0.49\columnwidth}
         \centering
           \includegraphics[width=1\columnwidth]{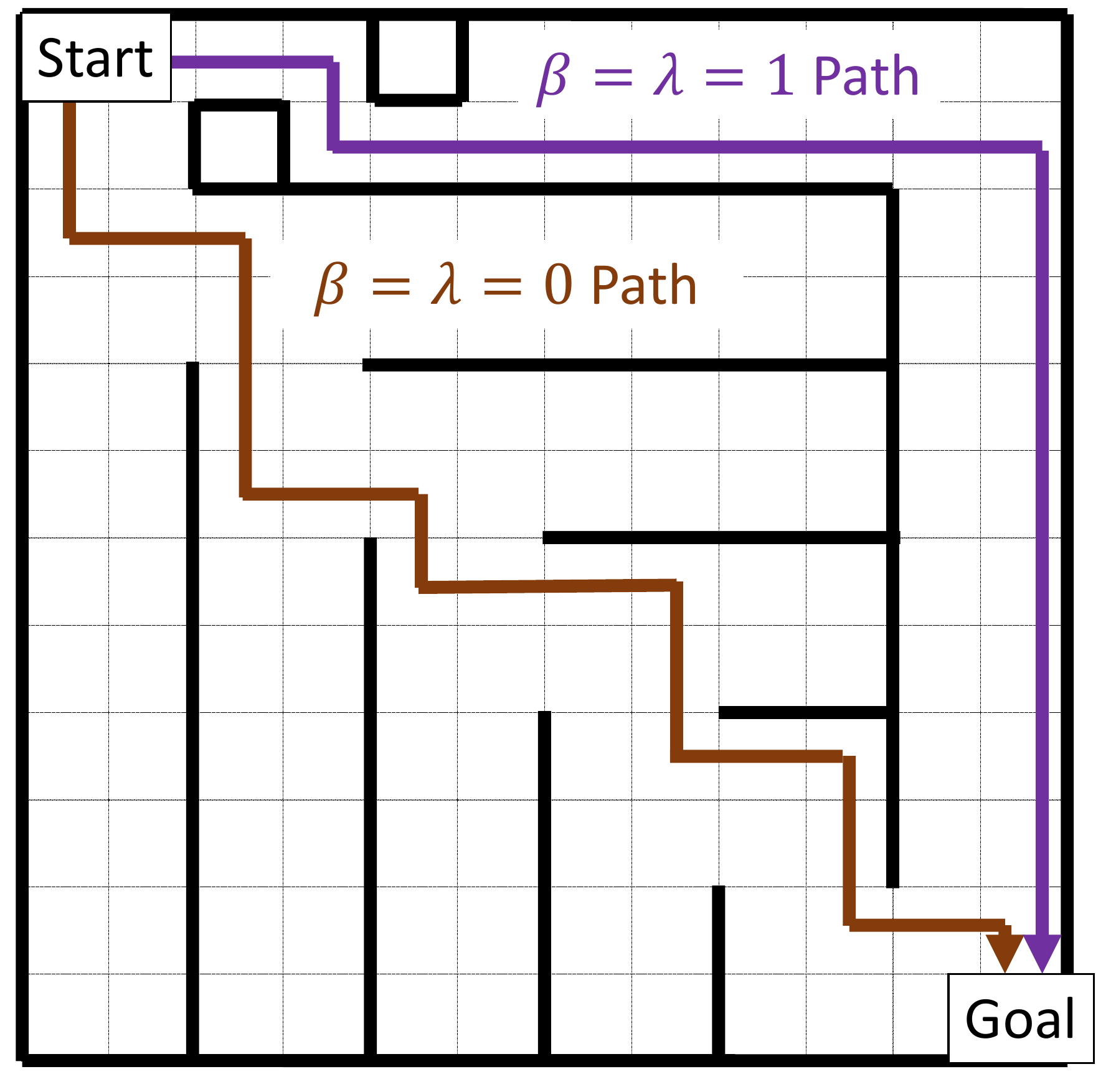}
         \caption{}
         \label{fig:path1}
     \end{subfigure}
     \begin{subfigure}[h!]{0.49\columnwidth}
         \centering
          \includegraphics[width=1\columnwidth]{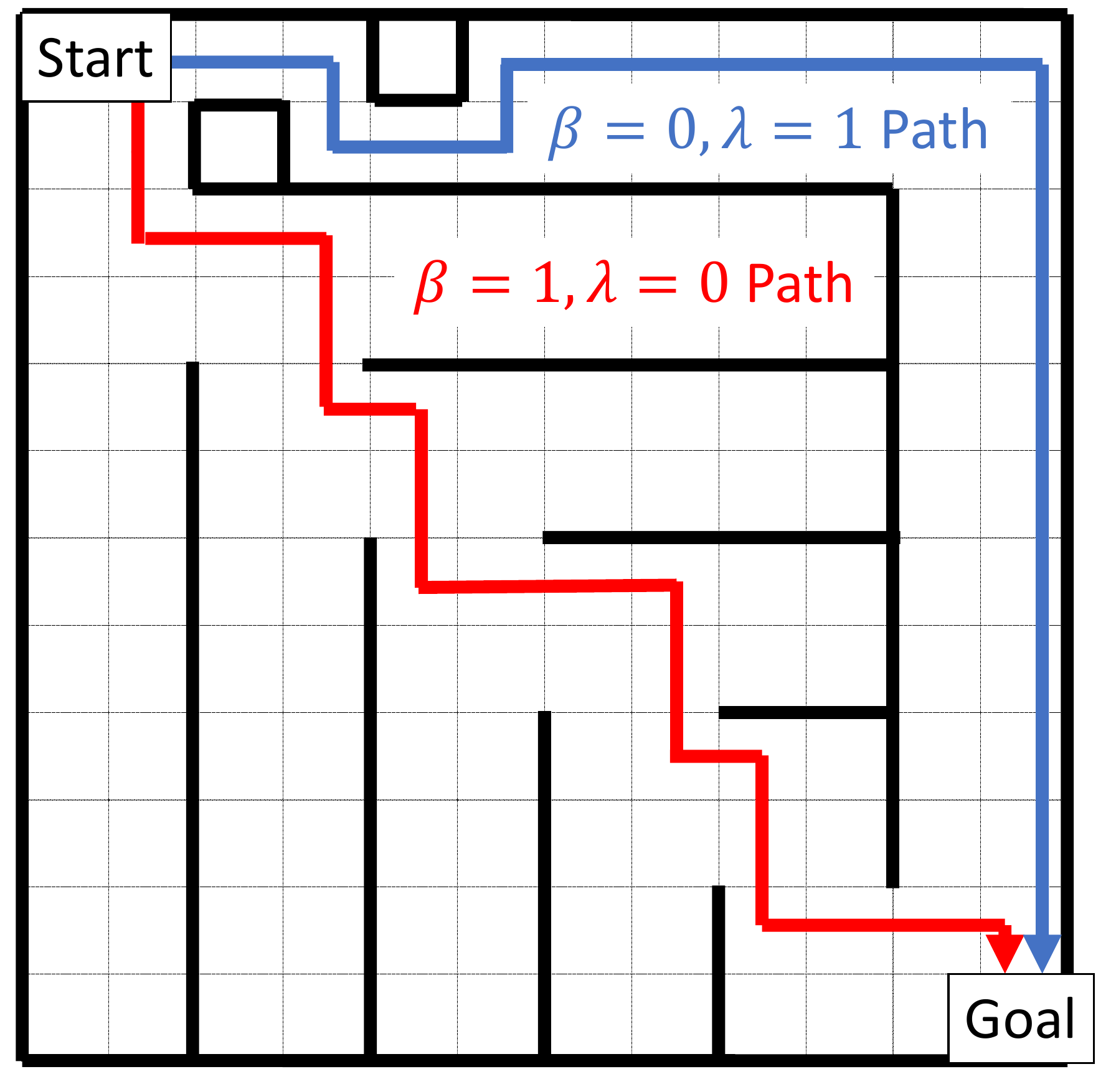}
         \caption{}
         \label{fig:path2}
     \end{subfigure}
    \caption{Example realizations of ERPOMDP \eqref{eq:er} agent with wall sensor moving from start cell (unknown to agent, top-left in these realizations) to goal (walls bold black): (a) $\beta = \lambda \in \{0,1\}$ (b) $\beta = 1$ \& $\lambda = 0$, and $\beta = 0$ \& $\lambda = 1$.}
    \label{fig:path}
    %\vspace{-0.5cm}
\end{figure}

Consider an agent in the grid shown in Fig.\ \ref{fig:path}, that seeks to move to (and stay in) a known goal location from an unknown starting location { (distributed uniformly over the grid such that the initial state pmf $\rho$ is uniform)}, whilst actively localizing itself so as to enable its path to the goal to be estimated for the purpose of later being retraced or communicated.
Each cell in the grid is a state in the agent's state space $\mathcal{X} = \{1, \ldots, 144\}$ (enumerated top-to-bottom, left-to-right).
The agent has five possible control actions $\mathcal{U} = \{1, \ldots, 5\}$, corresponding to moving one cell in each of the four compass directions, or staying still (all with probability $1$).
There are internal and external walls (bold black lines in Fig.\ \ref{fig:path}) that block movement, with the agent staying still if it attempts to move into them.
The agent receives measurements $\mathcal{Y} = \{1,\ldots,16\}$ corresponding to whether or not a wall is immediately adjacent to its current cell in each of the four compass directions. 
The agent detects a wall when it is present (resp.\ not present) with probability $1$ (resp.\ $0.2$).
A simplified version of this example was previously considered in \cite{Tanaka2021} for MDPs.% and in \cite{Molloy2022} (for POMDPs with more simplistic direction-less wall measurements).

{
We examine the ability of the agent to move to the goal and ensure estimation of its path by solving \eqref{eq:er} with either $\beta = \lambda = 0$ (corresponding to a standard POMDP without any regularization), $\beta = \lambda = 1$ (corresponding to joint-entropy regularization), $\beta = 1$ and $\lambda = 0$ (corresponding to only smoother-entropy regularization), and $\beta = 0$ and $\lambda = 1$ (corresponding to only input-output-entropy regularization).
In all cases, $\gamma = 0.99$ and the goal objective is encoded via the cost $c(x, u) = 1_{\{x \neq 144\}}$ for all $x \in \mathcal{X}$ and $u \in \mathcal{U}$.

We use SARSOP \cite{Kurniawati2008} to solve \eqref{eq:er} via the reformulation in Theorem \ref{theorem:jer} when $\beta = \lambda \in \{0,1\}$, and via the reformulation in Theorem \ref{theorem:MDP} when $\beta \neq \lambda$.
For $\beta \neq \lambda$, we construct a PWLC approximation of $G$ in \eqref{eq:er_mdp} using a set $\Xi$ containing the middle of the simplex $\Delta$ and points near the vertices with values in their largest element of $0.857$ and $0.001$ in their other $143$ elements.
For $\beta = \lambda$, we avoid PWLC approximations since the cost function $L$ in Theorem \ref{theorem:jer} is linear.
From Table \ref{tbl:navigation}, we see that the time taken to compute policies requiring PWLC approximations (i.e., policies with $\beta \neq \lambda$) is much greater than the time required to compute the standard POMDP policy with $\beta = \lambda = 0$, and the ERPOMDP policy with $\beta = \lambda = 1$.
} %able to directly optimize the joint trajectory entropy $H_\mu(X^T, Y^T, U^{T-1})$ without introducing any approximations, extra computational steps, or software modifications.
%As a result, the dimensions of the state and observation spaces in this example are able to far exceed those of other similar examples in \cite[Section 6]{Fehr2018}, \cite[Chapter 8]{Krishnamurthy2016}, and \cite{Molloy2021a,Molloy2021b,Molloy2022}.

% 
% , 
% , 
% , 
% , 
% , $P(\hat{X}^T \neq X^T)$

\subsection{Simulation Results}

The results of $1000$ Monte Carlo simulations of each policy over $T = 100$ time steps (the mean of $T$) are summarized in Table \ref{tbl:navigation}.
For each policy, we report several active state estimation criteria including: the total (undiscounted) cost associated with not being in the goal state (i.e., the \emph{Goal Cost}) $\sum_{k = 0}^T E[c(X_k, U_k)]$; the input-output entropy; the smoother entropy; the joint entropy; the sum of belief entropies $\sum_{k = 0}^T H_\mu(X_k | Y^k, U^{k-1})$; and, the probability of error in maximum \emph{a posteriori} estimates of the trajectory $X^T$ (\emph{Traj. MAP Error Prob.}) computed via the Viterbi algorithm \cite[Section 3.5.3]{Krishnamurthy2016}.
{ Example state realizations with the agent starting in the top-left cell are shown in Fig.~\ref{fig:path}.
%Fig.\ \ref{fig:path} shows the (instantaneous) entropy of the belief state $\pi_k$ for these realizations (i.e., $H(X_k | y^k, u^{k-1})$) with its expected sum being \emph{Sum Bel.\ Ents.}).

Table \ref{tbl:navigation} shows that the standard POMDP policy ($\beta = \lambda = 0$) results in the lowest goal cost, which is unsurprising since it only explicitly minimizes the (discounted) cost of the agent not being in the goal state.
The smoother entropy, input-output entropy, and joint entropy are all significantly less (better) when they are regularized via selection of $\beta = 1$ and $\lambda = 0$, $\lambda = 1$ and $\beta = 0$, or both $\beta = \lambda = 1$, respectively (at the expense of a small increase in the \emph{Goal Cost}).
Table \ref{tbl:navigation} thus highlights that the agent resolves its uncertainty and reduces the memory required to store its measurements more effectively with versions of ERPOMDP policies with nonzero $\beta$ and $\lambda$ than with the standard POMDP policy (with $\beta = \lambda = 0$).

%, sum of belief entropies, and \emph{Traj. MAP Error Prob.} in Table \ref{tbl:navigation} 
As illustrated in Fig.\ \ref{fig:path}, the ERPOMDP policies with $\beta = \lambda = 1$, and $\beta = 0$ and $\lambda = 1$, reduce the joint entropy and input-output entropy most effectively by moving the agent through the cells along the walls since these yield easily compressible observation sequences (a single wall in the same relative direction).
In contrast, the cells in the center through which the standard POMDP policy (with $\beta = \lambda = 0$) and the ERPOMDP policy with $\beta = 1$ and $\lambda = 0$ move the agent yield more variable (higher entropy) observations since they are surrounded by $0$, $1$, or $2$ walls.
By initially moving the agent East from its starting cell and then through the center, the ERPOMDP policy with $\beta = 1$ and $\lambda = 0$ is able to achieve the minimum smoother entropy.
However, in this example, the differences in smoother entropies between policies with $\beta = 0$ and $\beta = 1$ is much smaller compared to the differences in input-output entropies between policies with $\lambda = 0$ and $\lambda = 1$.
Thus, the ERPOMDP policy involving joint-entropy regularization via $\beta = \lambda = 1$ achieves close to the best performance across the criteria (at the slight expense of the \emph{Goal Cost}) whilst avoiding PWLC approximations, highlighting the value of Theorem \ref{theorem:jer}.
Clearly, finer trade-offs between the \emph{Goal Cost} and smoother or input-output entropies can be obtained by selecting $\beta$ and $\lambda$ independently and using Theorem \ref{theorem:MDP}, but at considerable computational cost.}

% \usepackage{booktabs}
% \usepackage{multirow}
\begin{table}[t!]
\centering
\caption{Monte Carlo Simulation Results (best values in bold).
Computational times for an M1 2020 Apple MacBook Air.}
\label{tbl:navigation}
\begin{tabular}{@{}lcccc@{}}
\toprule
\multicolumn{1}{c}{\multirow{3}{*}{\textbf{\begin{tabular}[c]{@{}c@{}}Performance\\ Criteria\end{tabular}}}} & \multicolumn{4}{c}{\textbf{ERPOMDP \eqref{eq:er} Policy Parameters}} \\ \cmidrule(l){2-5} 
\multicolumn{1}{c}{}                                                                                    & $\beta = 0$ & $\beta = 1$ & $\beta = 0$ & $\beta = 1$\\
\multicolumn{1}{c}{}                                                                                    & $\lambda = 0$ & $\lambda = 0$ & $\lambda = 1$ & $\lambda = 1$\\
\midrule
\multicolumn{1}{l|}{\textbf{Goal Cost}}                                 & \textbf{23.0}  & 25.2           & 25.9           & 26.0 \\
\multicolumn{1}{l|}{\textbf{Input-Output Entropy}}                      & 120.2          & 122.0          & \textbf{114.9} & \textbf{114.9} \\
\multicolumn{1}{l|}{\textbf{Smoother Entropy}}                          & 1.47           & \textbf{0.41}  & 0.60           & 0.50\\
\multicolumn{1}{l|}{\textbf{Joint Entropy}}                             & 121.7          & 122.4          & 115.5          & \textbf{115.4}\\
\multicolumn{1}{l|}{\textbf{Sum of Belief Entropies}}                        & 21.9           & 17.3           & \textbf{16.0}  & \textbf{16.0}\\
\multicolumn{1}{l|}{\textbf{Traj. MAP Error Prob.}}                     & 0.01           & \textbf{0.00}  & 0.01           & \textbf{0.00}\\
\multicolumn{1}{l|}{\textbf{Time to Compute Policy (s)}}                   & \textbf{0.21}            & 6563           & 964            & 0.66 \\\bottomrule
\end{tabular}
\end{table}

\section{Conclusion}
\label{sec:conclusion}
We propose ERPOMDPs, show that they admit PWLC approximate solutions, and discuss their relevance to active state estimation and rate-cost trade-offs.
Surprisingly, ERPOMDPs admit exact solutions when regularizing by the joint entropy of the states, observations, and controls, which constitutes a novel, tractable formulation of active state estimation.
%We present simulations in which ERPOMDPs reduce the probability of error in state trajectory estimates and the entropy of observations compared to standard POMDPs.

%\section*{Appendix}
%\label{sec:appendix}

% \section*{Acknowledgement}

%% References section
\bibliographystyle{IEEEtran}
%\balance
\bibliography{IEEEabrv,Library}

\end{document}